\documentclass[final]{svjour2}
\usepackage{graphicx}
\usepackage{rotating}
\usepackage{amssymb}
\usepackage{mathptmx}
\usepackage[numbers]{natbib}
\makeatletter
\journalname{Journal of Low Temperature Physics}

\bibpunct{}{}{,}{s}{}{,}

\begin{document}

\newcommand{\hdblarrow}{H\makebox[0.9ex][l]{$\downdownarrows$}-}

\title{Absolute Energy Calibration of X-ray TESs with 0.04 eV Uncertainty at 6.4 keV in a Hadron-Beam Environment}

\author{H.~Tatsuno \and W.B.~Doriese \and D.A.~Bennett \and C.~Curceanu 
\and J.W.~Fowler \and J.~Gard 
\and F.P.~Gustafsson \and T.~Hashimoto \and R.S.~Hayano 
\and J.P.~Hays-Wehle \and G.C.~Hilton \and M.~Iliescu 
\and S.~Ishimoto \and K.~Itahashi \and M.~Iwasaki 
\and K.~Kuwabara \and Y.~Ma \and J.~Marton \and H.~Noda 
\and G.C.~O'Neil \and S.~Okada \and H.~Outa \and C.D.~Reintsema 
\and M.~Sato \and D.R.~Schmidt \and H.~Shi \and K.~Suzuki 
\and T.~Suzuki \and J.~Uhlig \and J.N.~Ullom 
\and E.~Widmann \and S.~Yamada \and J.~Zmeskal \and D.S.~Swetz}

\institute{H.~Tatsuno \and S.~Ishimoto\\
High Energy Accelerator Research Organization, KEK, Ibaraki 305-0801, Japan\\
\email{hideyuki.tatsuno@gmail.com}\\
\\
D.A.~Bennett \and  W.B.~Doriese \and  J.W.~Fowler \and J.~Gard \and
J.P.~Hays-Wehle \and G.C.~Hilton \and G.C.~O'Neil \and C.D.~Reintsema \and
D.R.~Schmidt \and D.S.~Swetz \and J.N.~Ullom\\
National Institute of Standards and Technology, Boulder, CO 80305, USA\\
\\
C.~Curceanu \and M.~Iliescu \and H.~Shi\\
Laboratori Nazionali di Frascati, INFN, I-00044 Frascati, Italy\\
\\
F.P.~Gustafsson \and J.~Uhlig\\
Department of Chemical Physics, Lund University, Lund, 221 00, Sweden\\
\\
T.~Hashimoto \and K.~Itahashi \and M.~Iwasaki \and Y.~Ma \and H.~Noda \and
S.~Okada \and H.~Outa \and M.~Sato\\
RIKEN Nishina Center, RIKEN, Wako, 351-0198, Japan\\
\\
R.S.~Hayano \and T.~Suzuki\\
Department of Physics, The University of Tokyo, Tokyo, 113-0033, Japan\\
\\
K.~Kuwabara \and S.~Yamada\\
Department of Physics, Tokyo Metropolitan University, Tokyo, 192-0397, Japan\\
\\
J.~Marton \and K.~Suzuki \and E.~Widmann \and J.~Zmeskal\\
Stefan-Meyer-Institut f\"{u}r Subatomare Physik, Vienna, A-1090, Austria}

\maketitle

\begin{abstract}

A performance evaluation of superconducting transition-edge sensors (TESs) in the environment of a 
pion beam line at a particle accelerator is presented. Averaged across the 209 functioning sensors in 
the array, the achieved energy resolution is 5.2 eV FWHM at Co $K_{\alpha}$ (6.9 keV) when the pion 
beam is off and 7.3 eV at a beam rate of 1.45 MHz. Absolute energy uncertainty of $\pm$0.04 eV 
is demonstrated for Fe $K_{\alpha}$ (6.4 keV) with \textit{in-situ} energy calibration obtained from 
other nearby known x-ray lines. To achieve this small uncertainty, it is essential to consider the 
non-Gaussian energy response of the TESs and thermal cross-talk pile-up effects due to 
charged-particle hits in the silicon substrate of the TES array.


\end{abstract}

\section {Introduction}
Recently, superconducting transition-edge sensors (TESs) have achieved eV-scale energy resolution in hard x-ray spectroscopy\cite{Ullom2005,Smith2012}. Due to its combination of energy resolution and collecting efficiency, an array of TESs is a powerful tool to determine the central energy and width of a low-intensity characteristic x-ray line. TES arrays are thus well-matched to the precision measurement of the strong-interaction shift and width of x-ray lines produced by hadronic atoms. 

A hadronic atom is a Coulomb-bound system of a negatively charged hadron (e.g., $\pi^-$, $K^-$, $\bar{p}$, $\Sigma^-$, $\Xi^-$), electrons, and a nucleus, enabling study of the fundamental strong nuclear force in the low-energy limit. In such a system, the increase in the reduced mass of the atom shifts all x-ray transitions from their standard-atomic values to much higher energies (e.g., keV x-rays for light atoms). A hadronic atom is created by stopping a hadron beam in a target. The slowed (non-relativistic) hadron is captured into a highly excited state and then cascades down to lower energy states via Auger and radiative processes. Finally, the hadron is absorbed by the nucleus in a process mediated by the strong force. The effects of the strong interaction appear in the lowest orbital as an energy shift from the electromagnetic energy level and a broadening due to the absorption\cite{Batty1997}. The shift and width are unique probes of the strong interaction.

TES-based kaonic-atom x-ray spectroscopy will explore the strong interaction of the $K^-$ with various nuclei. The depth of the ``optical potential'' that defines the $K^-$-nucleus interaction is still not determined due to the lack of precision in spectroscopic results. Presently, both the `deep-potential' phenomenological approach\cite{Batty1997,Friedman2011} and the `shallow-potential' chiral-unitary model\cite{Hirenzaki2000} can explain the measured kaonic-atom strong-interaction shifts and widths. To solve the `deep or shallow problem,' precision of $\pm$0.2 eV or better is required for the measurement of the kaonic helium 3d--2p x-ray energy (6.4 keV). The current best precision for the strong-interaction shift measurements of this line is $\pm$2 eV\cite{Okada2012}.

To prove the viability of the TES-based measurement technique and to evaluate the performance of TESs in the charged particle environment of an accelerator beam line, we performed a demonstration experiment at the $\pi$M1 pion beam line at the Paul Scherrer Institute (PSI; Villigen, Switzerland).

\section{Demonstration experiment at the PSI $\pi$M1 beam line}
\begin{figure}[t]
\begin{center}
 \includegraphics[width=0.75 \linewidth]{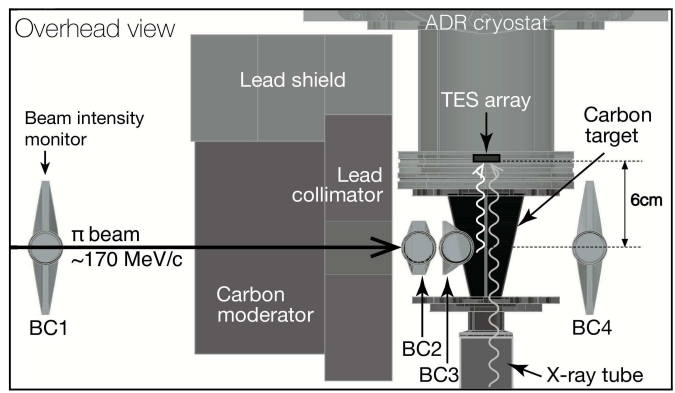}
\end{center}
\caption{Overhead view of the setup for the $\pi$-$^{12}$C x-ray measurement at the 
         PSI $\pi$M1 beam line. The BCs are beam-line counters made with plastic 
         scintillators and photo multiplier tubes to trigger the stopped pions. 
         The distance from the TES array to the beam center is 6 cm.}
\label{Fig:PSI_setup}
\end{figure}

Figure \ref{Fig:PSI_setup} shows the experimental setup at the PSI $\pi$M1 beam line. Incident 170 MeV/$c$ pions are slowed down in carbon moderators, and stopped in a carbon sample target. Pionic-carbon ($\pi$-$^{12}$C) x-rays are detected by the TES spectrometer. To provide calibration x-rays constantly, Bremsstrahlung emission from an x-ray tube induces fluorescence in high purity Cr and Co foil targets. These fluorescence x-rays pass through a hole in the carbon sample target, allowing \textit{in-situ} energy calibration of the TESs during the measurements. The x-rays enter through the cryostat's vacuum window (150-$\mu$m thick Be) and pass through three layers of IR-blocking filters (each 5-$\mu$m thick Al) at three temperature stages (50 K, 3 K, and 50 mK) before reaching the 240-pixel TES array\cite{Ullom2014}. Each TES consists of a superconducting bilayer of thin Mo and Cu films with an additional 4-$\mu$m thick Bi absorber ($\sim 80$\% absorption efficiency for 6.4 keV x-rays), and has an active area of 320 $\mu$m $\times$ 305 $\mu$m set by a gold-coated Si collimator placed in front of the array, thus the total active area of the array is about 23 mm$^2$. The TES array is cooled with a pulse-tube-backed adiabatic demagnetization refrigerator (ADR)\cite{Bennett2012}. The ADR's bath temperature is regulated to 75 mK $\pm7$ $\mu$K (r.m.s). The TES pixels are then electrically biased to their superconducting critical temperature of  $T_{\rm C} \sim 100$ mK. The TESs' time-division-multiplexing readout system samples the current signal of the 240 sensors through 8 SQUID columns\cite{Reintsema2003,Reintsema2009}. The effective sampling rate of each detector is 104 kHz. The recorded number of samples is 1024 (= 9.83 ms) for each x-ray pulse. 

\section{Energy calibration}
\begin{figure}[bt]
 \begin{center}
   \begin{minipage}{7.1cm}
     \includegraphics[width=1.0 \linewidth]{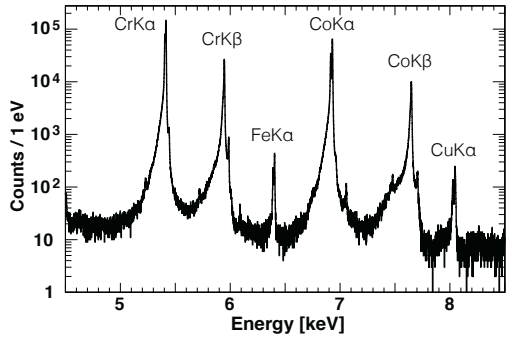}
   \end{minipage}
   \begin{minipage}{4.2cm}
     \caption{The calibration x-ray energy spectrum summed over 209 pixels 
     with the pion beam off. The Cr and Co fluorescence lines provide the 
     energy calibration. The lower-intensity x-ray lines, Fe $K_\alpha$ (6.4 keV) 
     and Cu $K_\alpha$ (8.0 keV), are backgrounds from surrounding metal parts 
     and can be used to verify the energy calibration. The achieved energy resolution 
     is 5.2 eV FWHM at Co $K_\alpha$ at a count rate of 4.4 Hz/pixel.}
    \label{Fig:CrCoCalibration}
   \end{minipage}
 \end{center}
\end{figure}

The x-ray pulse height is calculated with the optimal filter technique\cite{Szymkowiak1993}. The optimal filter is constructed from the average pulse shape and the noise power spectral density. The pulse-height and energy relation of each sensor is determined with Cr $K_\alpha$, Cr $K_\beta$, Co $K_\alpha$, and Co $K_\beta$ lines. The peak positions of pulse-height histogram are fixed at the known energies from H\"{o}lzer \textit{et al}.\cite{Holzer1997}, and then a cubic-spline interpolation curve is formed in the energy region of interest. Figure \ref{Fig:CrCoCalibration} shows an x-ray energy spectrum measured with the TESs installed at the $\pi$M1 beam line with the pion beam off. The lower-intensity x-ray lines, Fe $K_\alpha$ (6.4 keV) and Cu $K_\alpha$ (8.0 keV), originate from the stainless-steel fittings that surround the detector and the Cu base of the detector package, respectively. The energy resolution of combined functioning 209 sensors is 5.2 eV FWHM at Co $K_\alpha$ at a count rate of 4.4 Hz/pixel.

\begin{figure}[thb]
\begin{center}
 \includegraphics[width=1.00 \linewidth]{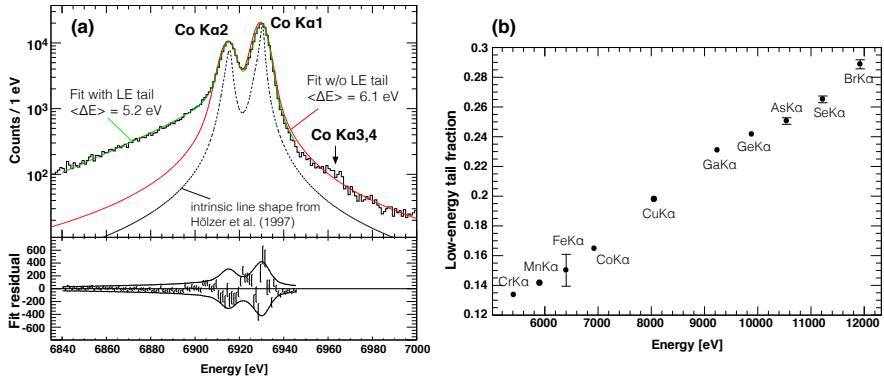}
\end{center}
\caption{({\bf a}) Co $K_{\alpha}$ spectrum with the pion beam off. The \textit{green} curve shows 
         the fit with the LE tail; its residual is shown at \textit{bottom} with $\pm3\sigma$ curves.
         The fit range is selected to avoid the Co $K_{\alpha3}$ and $K_{\alpha4}$ satellite peaks.
         The \textit{red} curve shows the fit without considering the LE tail. 
         The \textit{broken} curve shows the intrinsic line shape from 
         H\"{o}lzer \textit{et al.}\cite{Holzer1997}. ({\bf b}) The measured low-energy 
         tail fraction for several elements with statistical errors 
         (some error bars are too small to see). 
         The tail fractions of Mn, Cu, Ga, Ge, As, Se, and Br are measured with another 
         calibration target. (Color figure online.)}
\label{Fig:LEtail}
\end{figure}

The response function of the TESs must be well understood for accurate energy calibration. A clear non-Gaussian response is observed in the spectrum, and is empirically represented as a low-energy (LE) tail function, the convolution of an exponential and a Gaussian:
\begin{equation}
 {\rm LE~tail}(E,\sigma,\beta) = A~{\rm exp} \left( \frac{E-E_0}{\beta} \right)
 ~{\rm Erfc} \left( \frac{E - E_0}{\sqrt{2}\sigma} + \frac{\sigma}{\sqrt{2}\beta} \right),
\end{equation}
where $E_0$ is the x-ray energy, $\sigma$ is the Gaussian sigma, $\beta$ is the slope parameter of the exponential, and $A$ is the intensity factor which also defines the tail fraction. Erfc is the complementary error function. The physical source of this LE tail is not fully understood at this time. Preliminary evidence suggests it originates in the thermally evaporated Bi absorbers. The grain structure of Bi may trap the heat in its lattice with a thermal decay time that is much longer than the thermal response time of the TES itself.
The LE-tail function is well fitted as shown in Fig.~\ref{Fig:LEtail}(a). The LE-tail fraction is 16.5\% for Co $K_{\alpha}$ and $\beta$ is 30.0 eV. Fitting with vs. without the LE tail results in a shift in the peak position of 0.5 eV, so inclusion of the tail is critical to reach the precision of $\pm$0.2 eV required for this experiment and future kaonic-atom spectroscopy. The LE-tail fraction for several elements is measured as shown in Fig.~\ref{Fig:LEtail}(b). It increases linearly with the x-ray energy. This relation constrains the LE-tail fraction in lines that are too dim to allow the tail fraction to be measured directly.

\begin{figure}[h]
 \begin{center}
  \includegraphics[width=1.0 \linewidth]{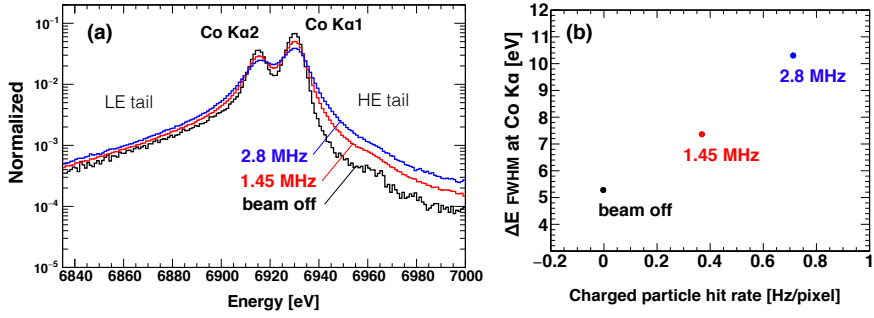}
 \end{center}
 \caption{({\bf a}) Spectra of Co $K_{\alpha}$ x-rays with different pion-beam 
          rates (particles per second). The \textit{black} is beam off, 
          the \textit{red} is 1.45 MHz, and the \textit{blue} is 2.8 MHz. 
          ({\bf b}) Degradation in the Gaussian energy resolution as a function of the pion-beam 
          rate. (Color figure online.)}
\label{Fig:Beamon}
\end{figure}

The pion-beam-on spectrum of Co $K_\alpha$ has additional low-energy and high-energy (HE) components as shown in Fig.~\ref{Fig:Beamon}(a). In addition, Fig.~\ref{Fig:Beamon}(b) shows the degradation in the Gaussian component of the energy resolution as a function of beam rate (particles per second). These beam-correlated effects are explained by the pile-up of thermal cross-talk pulses. When a high-energy charged particle penetrates the 275-$\mu$m thick Si substrate of the TES array, about 100 keV of energy is deposited. This energy thermalizes rapidly and heats the entire array, and in so doing, creates cross-talk pulses in all TESs. Typical optimal filtering for a piled-up pulse gives a poor estimation of the x-ray energy, because the piled-up pulse shape does not match the average pulse shape. The estimated energy can shift lower or higher depending on where in time the cross-talk pulse coincides relative to the x-ray pulse. Some of piled-up events are identified and cut, which does yield a moderate improvement in the measured spectrum. Though, it is difficult to distinguish the piled-up pulses close to each other in time. Therefore, an additional HE-tail function same as Eq.~(1) is required to fit the spectrum with the pion beam on. The fitted HE-tail fraction is 8.3\%.

\section{Absolute Energy Determination}
\begin{figure}[tb]
\begin{center}
 \begin{minipage}{7.3cm}
  \includegraphics[width=1.0 \linewidth]{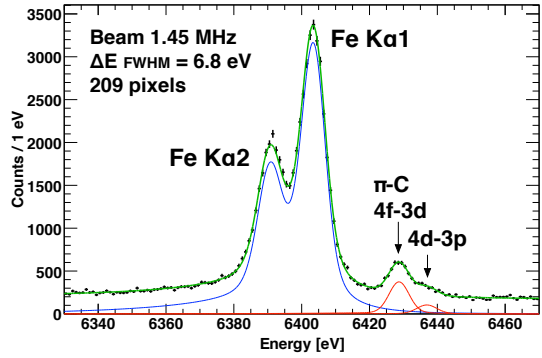}
 \end{minipage}
 \begin{minipage}{4.2cm}
  \caption{The measured x-ray spectrum of Fe $K_{\alpha}$ and the $\pi$-$^{12}$C 4--3 transitions 
           summed over 209 pixels at the 1.45 MHz beam rate. The \textit{green} curve shows 
           the fitting to the summed spectrum and a linear background. The \textit{blue} curve is the 
           component of the fit due to only the Fe $K_{\alpha}$ x-rays with LE and HE tails, 
           and the \textit{red} curves are due to the individual $\pi$-$^{12}$C lines with tails. 
           The energy resolution is 6.8 eV FWHM with the pion beam on. (Color figure online.)}
  \label{Fig:PiCResults}
 \end{minipage}
\end{center}
\end{figure}

The background Fe $K_\alpha$ x-rays provide a useful test of the energy calibration very near the energy of the $\pi$-$^{12}$C line of interest. Figure \ref{Fig:PiCResults} shows the measured x-ray spectrum of the Fe $K_{\alpha}$ and the $\pi$-$^{12}$C 4--3 transitions (to be discussed in a Letter) with 209 TES pixels at the 1.45 MHz beam rate. The \textit{in-situ} energy calibration is re-calculated every 2 hours to mitigate gain drift of TESs. The absolute energy of Fe $K_{\alpha}$ is consistent with the value of H\"{o}lzer \textit{et al.}\cite{Holzer1997} within $\pm0.03$ eV statistical error $\sigma_{\rm stat}$ and $\pm0.03$ eV systematic uncertainty $\sigma_{\rm syst}$. The main uncertainty source is the ambiguity of LE-tail (absorber and thermal cross-talk origins) and HE-tail parameters as interpolated from the fit values for Cr $K_\alpha$ and Co $K_\alpha$ ($\pm0.02$ eV). The total estimated uncertainty of the absolute energy is $\pm0.04$ eV (=$\sqrt{\sigma_{\rm stat}^2 + \sigma_{\rm syst}^2}$). This precision is about 10 times better than the measurement technique with silicon drift detectors\cite{Okada2012}, and meets our precision goal for the future measurement of kaonic-helium $3d$--$2p$ x-ray energy.

\section{Conclusion}
We have evaluated the performance of a 240-pixel TES array at the PSI $\pi$M1 beam line, and achieved absolute energy uncertainty of $\pm0.04$ eV for Fe $K_{\alpha}$ in the charged-particle beam environment. Charged-particle impacts on the TES array are investigated for the first time: the piled-up thermal cross-talk pulses degrade the Gaussian energy resolution and create low-energy and high-energy tails in the x-ray spectrum. Even in the absence of the charged-particle beam, the TES energy response also shows a LE tail, which could be caused by the grain structure of the Bi absorber. While these LE and HE tails add complexity to the analysis of the data, they do not prevent the TESs from achieving absolute energy calibration at a level necessary for TES-based kaoinic-atom spectroscopy to solve the $K^-$-nucleus potential depth problem. In future experiments, better shielding will limit the LE and HE tails from charged-particle hits, and the LE-tail response will be eliminated via a different absorber fabrication process.

\begin{acknowledgements}
We thank K.~Deiters and the PSI staff for the support and the beam-line coordination and operation. We thank as well the members in the NIST Quantum Sensors Project for the efforts and discussions to realize this challenging application. This work was partly supported by the Grants-in-Aid for Scientific Research from MEXT and JSPS (Nos.~25105514, 26707014, 24105003 and 15H05438), the strategic young researcher overseas visits program for accelerating brain circulation by JSPS (No. R2509), and the NIST Innovations in Measurement Science Program. Submission of an agency of the U.S.~Government; not subject to copyright.
\end{acknowledgements}

\end{document}